\documentclass[a4paper,fleqn]{cas-dc}
\usepackage[numbers,sort&compress]{natbib}
\usepackage[switch]{lineno}
\begin{document}
\let\WriteBookmarks\relax
\def\floatpagepagefraction{1}
\def\textpagefraction{.001}
\shorttitle{Infinite dilute activity coefficients for Ga-X (X= In, Tl) and component thermodynamic activities of liquid Ga-In-Tl alloys}
\shortauthors{Oshakuade and Awe}
\title [mode = title]{Computation of infinite dilute activity coefficients for Ga-X (X= In, Tl) and thermodynamic activities of all components in liquid Ga-In-Tl alloys}                      
\author{O.M. Oshakuade}[type=author,orcid=0000-0001-9971-1213]
\cormark[1]
\ead{om.oshakuade@ui.edu.ng}
\address{Department of Physics, University of Ibadan, Ibadan, Nigeria}
\author{O.E. Awe}
\cortext[cor1]{Corresponding author}

\begin{abstract}
The thermodynamic activities of all components in Ga-In-Tl system have been predicted at 1073, 1173 and 1273 K, using Molecular interaction volume model (MIVM). The infinite dilute activity coefficients for Ga-In and Ga-Tl binary subsystems, which were needed for the determination of thermodynamic activities of all components in Ga-In-Tl, have been predicted by using a method that is based on Complex formation model for liquid alloys. The computed thermodynamic activities of Ga in Ga-In-Tl were observed to satisfactorily agree with the available experimental data when the newly computed coefficients were applied in MIVM. The satisfactory prediction of the activity of Ga led to the prediction of the activities of the remaining two components (In and Tl). Iso-activities of all components (Ga, In and Tl) were plotted, and they reveal the dependence of the nature of chemical short range order in Ga-In-Tl system on composition.
\end{abstract}


\maketitle 

\section{Introduction}
\label{intro}

Although, Ga-In-Tl has been identified as one of the Pb-free solder candidates, only the thermodynamic activities of Ga at 1073, 1173 and 1273 K have been studied experimentally \cite{pap01}. Since the knowledge of thermodynamic activities of all components in a system helps in understanding alloying of metals and process metallurgy, hence, the focus of this work is to use mathematical models to predict the non-existent thermodynamic activities of In and Tl in the Ga-In-Tl system at 1073, 1173 and 1273 K. 

Molecular interaction volume model (MIVM) \cite{pap02} is a theoretical tool that has been applied by many researchers \citep{pap03,pap04,pap05,pap06,pap07,pap08} to successfully predict thermodynamic activities  of components of liquid alloys. MIVM requires the knowledge of a few physical parameters for each alloy component, and infinite dilute activity coefficient pairs $(\gamma_i^\infty\text{ and }\gamma_j^\infty)$ for every binary sub-system. The infinite dilute activity coefficient of a binary alloy component is a thermodynamic property that reflects the property of a single atom of that component when it is completely surrounded by infinite atoms of the other alloy component \cite{pap09,pap10}. Also, $\gamma_i^\infty\text{ and }\gamma_j^\infty$ are relevant in scientific and engineering applications, such as: characterization of liquid mixture, estimation of solubility, and designing processes for separating dilute contaminants from water \cite{pap09,pap10}. Up till date, experimentally obtained $\gamma_i^\infty\text{ and }\gamma_j^\infty$ for Ga-In and Ga-Tl are non-existent, but fortunately, information on their integral Gibbs free energies of mixing  are available \cite{pap11,pap12}. Hence, a method for computing $\gamma_i^\infty\text{ and }\gamma_j^\infty$ using Complex formation model (CFM) \cite{pap13} can be used to obtain these values for Ga-In and Ga-Tl systems. 

The knowledge of the infinite dilute activity coefficients for every sub-system in Ga-In-Tl will aid the application of MIVM in this study. And good agreements between the computed and experimental activity of Ga will serve as premise to predict the unknown thermodynamic activities of In and Tl at 1073, 1173 and 1273 K.

The paper is laid out as follows: the next section gives details on the theoretical framework for computing infinite dilute activity coefficients and thermodynamic activities,  \autoref{results} contains results and discussions from the computations, while \autoref{conclusions} contains summary and conclusions.

\section{Theoretical framework}
\label{theory}

\subsection{Computation of infinite dilute activity coefficients}
\label{theory:gammainfinite}
Infinite dilute activity coefficients ($\gamma_i^\infty $ and $\gamma_j^\infty $) are usually obtained from experiments. However, the authors are not aware of any experimentally obtained $\gamma_i^\infty $ and $\gamma_j^\infty $ for Ga-In and Ga-Tl systems. Besides, experimental $G_M$ data exists for Ga-In and Ga-Tl at 623 and 973 K, respectively. With the knowledge of Gibbs free energies of mixing for the two systems, their $\gamma_i^\infty $ and $\gamma_j^\infty $ can be computed with the aid of Complex formation model (CFM) \cite{pap13,pap14}. The expression for $G_M$ in the Flory’s approximation of CFM is given by Eq. (\ref{eqn:1}) \cite{pap14}.
\begin{equation}
\label{eqn:1}
\begin{split}
& \resizebox{1.0\hsize}{!}{
$G_M=-n_3g +RT \left[n_1 \ln\frac{n_1}{N}+n_2 \ln\frac{n_2}{N}+n_3 \ln\frac{n_3\left(\mu+\nu \right)}{N} \right]$}\\
& \hspace{1cm} +\frac{n_1n_2\nu_{12}+n_1n_3\nu_{13}+n_2n_3\nu_{23}}{N} \\
\end{split}
\end{equation}
Where $R$ is molar gas constant, $T$ is system temperature, $N$ is the number of moles of system components (in their pure states), $\mu$ and $\nu$ are small integers obtainable from stoichiometric data, $n_1$, $n_2$ and $n_3$ are the number of moles of unassociated $i$ atoms, unassociated $j$ atoms, and $i_{\mu}j_{\nu}$ chemical complex, respectively, while $g$, $\nu_{12}$, $\nu_{13}$ and $\nu_{23}$ are fitted energy interaction parameters.

It is important to note that successful fit of experimental $G_M$ require obtaining $n_1$, $n_2$ and $n_3$ that will satisfy conservation equations: $n_1=N-\mu n_3$, $n_2=N-\nu n_3$ and the equilibrium condition $\partial G_M/\partial n_3$. A link between activity coefficients $(\gamma_i\text{, }\gamma_j)$ and the fitted interaction parameters is given by the following Eqs. (\ref{eqn:2}) and (\ref{eqn:3}).
\begin{equation}
\label{eqn:2}
\begin{split}
\ln \gamma_i &= \left.\ln\frac{n_1}{N_i}+\frac{N-n_1-n_2-n_3}{N}+\frac{1}{N R T}\right(n_2\nu_{12}  &\\
& \left.\hspace{1cm}+n_3\nu_{13}-\frac{n_1 n_2 \nu_{12}}{N}-\frac{n_1 n_3 \nu_{13}}{N}-\frac{n_2 n_3 \nu_{23}}{N}\right) &\\
\end{split}
\end{equation}
\begin{equation}
\label{eqn:3}
\begin{split}
\ln \gamma_j &= \left.\ln\frac{n_2}{N_j}+\frac{N-n_1-n_2-n_3}{N}+\frac{1}{N R T}\right(n_1\nu_{12}  &\\
& \left.\hspace{1cm}+ n_3\nu_{23}-\frac{n_1 n_2 \nu_{12}}{N}-\frac{n_1 n_3 \nu_{13}}{N}-\frac{n_2 n_3 \nu_{23}}{N}\right) &\\
\end{split}
\end{equation}
Where $N_i$ and $N_j$ are the number of moles of pure components $i$ and $j$, respectively. Therefore, $N_i+N_j=N$, and $c_k$ $(k = i \text{ or }j)$ is the molar fraction of component $k$ and its values is obtained from $c_k=N_k/N$ $(\text{where }k=i$\text{ or }$ j)$, and $c_i + c_j = 1$.

To obtain $\gamma_i^\infty$ and $\gamma_j^\infty$ from Eqs. (\ref{eqn:2}) and (\ref{eqn:3}), the following steps are important: (i.) fitting of the experimental $G_M$ of Ga-In and Ga-Tl in CFM using the energy interaction parameters \cite{pap14}, (ii.) applying the fitted energy interaction parameters in Eqs. (\ref{eqn:2}) and (\ref{eqn:3}) to obtain $\ln\gamma_i$ and $\ln\gamma_j$, (iii.) plotting the graphs of $\ln\gamma_i$ and $\ln\gamma_j$ versus molar concentrations $c_i$ and $c_j$, respectively, (iv.) using the polynomial fits to obtain the intercepts (which are $\ln\gamma_i^\infty$ and $\ln\gamma_j^\infty$), (v.) deducing the $\gamma_i^\infty$ and $\gamma_j^\infty$ for Ga-In and Ga-Tl from their natural logarithms and optimize by using different orders of polynomial fit.

\subsection{Molecular Interaction Volume Model (MIVM)}
\label{theory:mivm}
MIVM \cite{pap02} predicts thermodynamic activities $\left(a_is\right)$ from first-principles, it requires the knowledge of a few physical properties of the alloy components \cite{pap02,pap03,pap04,pap05,pap06,pap07,pap08}. It also requires pair potential parameters $\left(D_{ji}\right.$ and $\left.D_{ij}\right)$ which are derived from the infinite dilute activity coefficients $\left(\gamma_i^\infty\right.$ and $\left.\gamma_j^\infty\right)$ of each of the binary sub-systems \cite{pap02,pap13,pap15}. The natural logarithm of the activity coefficient of component $i$ in a $i-j$ binary or $i-j-k$ ternary system is defined in the following Eq. (\ref{eqn:4}) \cite{pap02,pap16}.
\begin{equation}
\label{eqn:4}
\begin{split}
& \ln \gamma_i=1+\ln \frac{V_{mi}}{\sum_{j=1}^\beta c_j V_{mj} D_{ji}}-\sum_{j=1}^\beta \frac{c_j V_{mi} D_{ij}}{\sum_{l=1}^\beta c_l V_{ml} D_{lj}} &\\
& \hspace{1cm}-\frac{1}{2} \left(\frac{Z_1 \sum_{j=1}^\beta c_j D_{ji} \ln D_{ji}}{\sum_{l=1}^\beta c_l D_{li}}+\sum_{j=1}^\beta\frac{Z_j c_j D_{ij}}{\sum_{l=1}^\beta c_l D_{lj}}\right. &\\
& \hspace{1cm} \left.\times\left(\ln D_{ij}-\frac{\sum_{l=1}^\beta c_l D_{lj} \ln D_{lj}}{\sum_{l=1}^\beta c_l D_{lj}}\right)\right) &
\end{split}
\end{equation}
Where $\beta$ is the number of components ($\beta = 2$ in binary and $\beta = 3$ in ternary systems), $Z_i$ is the coordination number of pure component $i$, while $V_{mi}$ is the molar volume of component $i$ at system temperature. The value of $Z_i$, is obtained from the following Eq. (\ref{eqn:5}) \cite{pap17}.
\begin{equation}
\label{eqn:5}
\resizebox{1.0\hsize}{!}{
$Z_i=\frac{4\sqrt{2\pi}}{3}\left(\frac{r_{mi}^3-r_{0i}^3}{r_{mi}-r_{0i}}\right) \frac{0.6022 r_{mi}}{V_{mi}} \exp \left(\frac{\Delta H_{mi}(T_{mi}-T)}{Z_c R\text{ } T\text{ } T_{mi}}\right)$}
\end{equation}
Where $\Delta H_{mi}$ and $\Delta T_{mi}$ are the melting enthalpy and melting temperature of component $i$, respectively, $Z_c$ is the close-packed coordination number of magnitude 12, $r_{0i}$ and $r_{mi}$ are the beginning and first peak values of radial distance distribution function near the melting point of liquid metal $i$, respectively. The pair-potential parameters $\left(D_{ji} \text{ and } D_{ij}\right)$ are defined in Eq. (\ref{eqn:6}) \cite{pap05}.
\begin{table*}[width=.9\textwidth,cols=10,pos=h]
\caption{Calculated $Z_i$ of each component and essential parameters related to its calculation.
{\label{tab:coord_nums}}}
\begin{tabular*}{\tblwidth}{@{} LLCCCCCCCC@{} }
\toprule
  Metal &  & ${\Delta H_{mi}\text{ }}^{a}$ &  &  &  \multicolumn{5}{c}{$Z_i \text{ } (T)\text{ }^{c}$}  \\   \cline{6-10}
  $i$ &  ${V_{mi}\text{ }}^{a} \text{ }(cm^3/mol)$ & $(kJ/mol)$ &   ${r_{mi}\text{ }}^{a,b}$   & ${r_{0i}\text{ }}^{a,b}$   & 623 K & 973 K &  1073 K & 1173 K & 1273 K \\
\midrule
  Ga & $11.40 [1 + 0.000092 (T - 303)]$ & 5.59 & 2.78 & 2.38 & 8.6754 & 8.1445 & 8.0313 & 7.9275 & 7.8311 \\
  In & $16.30 [1 + 0.000097 (T - 430)]$ & 3.26 & 3.14 & 2.70 & 9.5267 & $-$  & 8.9365 & 8.8327 & 8.7343 \\
  Tl & $18.00 [1 + 0.000115 (T - 576)]$ & 4.31 & 3.22 & 2.74 & $-$  & 8.9159 & 8.7825 & 8.6582 & 8.5413 \\
\bottomrule
\multicolumn{10}{l}{\footnotesize{$^a$ These parameters were obtained from \cite{pap18}}} \\ 
\multicolumn{10}{l}{\footnotesize{$^b$ Unit is $10^{-8}$ cm}}  \\ 
\multicolumn{10}{l}{\footnotesize{$^c$ The various values for different temperatures were obtained using Eq. (\ref{eqn:5})}}  \\ 
\end{tabular*}
\end{table*}

\begin{flalign}
\label{eqn:6}
& \text{  }D_{ji}=\exp\left[-\left(\frac{\epsilon_{ji}-\epsilon_{ii}}{k_B T}\right)\right] &
\end{flalign}
\begin{equation}
\label{eqn:7}
\begin{split}
& \ln\gamma_i^\infty =1-\ln\frac{V_{mj}D_{ji}}{V_{mi}}-\frac{V_{mi}D_{ij}}{V_{mj}} \hspace{2cm} &\\
& \hspace{1cm}-\frac{1}{2}\left(Z_i \ln D_{ji}+Z_j D_{ij} \ln D_{ij}\right) &\\
\end{split}
\end{equation}
\begin{equation}
\label{eqn:8}
\begin{aligned}
& \ln\gamma_j^\infty =1-\ln\frac{V_{mi}D_{ij}}{V_{mj}}-\frac{V_{mj}D_{ji}}{V_{mi}} \hspace{2cm} & \\
& \hspace{1cm}-\frac{1}{2}\left(Z_j \ln D_{ij}+Z_i D_{ji} \ln D_{ji}\right) & \\
\end{aligned}
\end{equation}
Where $\epsilon_{ii}$ and $\epsilon_{ji}$ are the $i-i$ and $i-j$ pair potential energies, respectively, and $k_B$ is Boltzmann’s constant.

Parameters $D_{ji}$ and $D_{ij}$ are obtained by solving Eqs. (\ref{eqn:7}) and (\ref{eqn:8}) (which requires the knowledge of $\gamma_i^\infty$ and $\gamma_j^\infty$), simultaneously. Equations (\ref{eqn:7}) and (\ref{eqn:8}) are solutions of Eq. (\ref{eqn:4}) for the binary system when $c_i \rightarrow 0$ and $c_j \rightarrow 0$, respectively \cite{pap05, pap16}. Then $D_{ji}$ at other temperatures were obtained by assuming ``$\epsilon_{ji} - \epsilon_{ii}$'' in Eq. (\ref{eqn:6}) to be temperature-independent, similarly, ``$\epsilon_{ij} - \epsilon_{jj}$'' was assumed to be temperature-independent to obtain $D_{ij}$ .

The $Z_i$ of Ga, In and Tl at different temperatures, and the parameters used in their computation are given in \autoref{tab:coord_nums} \cite{pap18}.

\section{Results and discussion}
\label{results}
To validate the results obtained from this work, the agreements between the existing and computed thermodynamic information were quantified by estimating Mean absolute percentage errors $(Er_t)$ and Standard errors $(Er_t^\ast)$, which are defined for a thermodynamic quantity, $t$, in Eqs. (\ref{eqn:9}) and (\ref{eqn:10}) \cite{pap05,pap19}, respectively.
\begin{flalign}
\label{eqn:9}
& \text{  }Er_t=\pm\frac{100}{\omega}\sum_{t=1}^\omega\left|\frac{d_{t,ex}-d_{t,pr}}{d_{t,ex}}\right| &
\end{flalign}
\begin{flalign}
\label{eqn:10}
& \text{  }Er_t^\ast=\pm\sqrt{\frac{1}{\omega}\sum_{t=1}^\omega {\left(d_{t,ex}-d_{t,pr}\right)}^2 } &
\end{flalign}
Where $d_{t,ex}$ and $d_{t,pr}$ are the existing and predicted values, respectively, while $\omega$ is the number of compared values.

\subsection{Infinite dilute activity coefficients}
\label{results:gammainfinite}
\begin{figure*}
    \includegraphics[width=\textwidth]{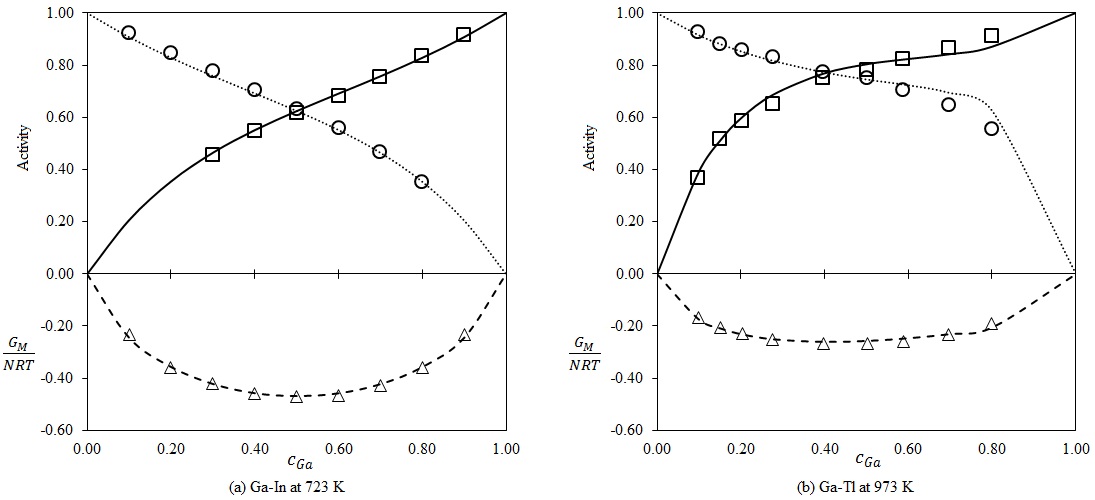}
    \caption{Composition dependence of $G_M/NRT$ and $a_is$. Triangles, squares and circles represent experimental $G_M/NRT$, $a_{Ga}$ and $a_{In}$ or $a_{Tl}$, respectively \cite{pap11,pap12}. Dashed lines, solid lines and dotted lines represent CFM-computed $G_M/NRT$, $a_{Ga}$ and $a_{In}$ or $a_{Tl}$, respectively.}
    \label{fig:cfidafm}
\end{figure*}

Following the method described in \autoref{theory:gammainfinite}, the experimental and CFM-fitted thermodynamic quantities for Ga-In and Ga-Tl systems are plotted in \autoref{fig:cfidafm}. The fitted interaction parameters used in CFM and the corresponding $Er_t$ for $G_M$ and $a_is$ are presented in \autoref{tab:cfm_data}. The plots in \autoref{fig:cfidafm} and the $Er_{G_M}$ and $Er_{a_is}$ values in \autoref{tab:cfm_data} show good agreement between the experimental and fitted data. The application of the interaction parameters in \autoref{tab:cfm_data} to Eqs. (\ref{eqn:2}) and (\ref{eqn:3}), followed by the quadratic polynomial fits of the resulting $\ln \gamma_i$ and $\ln \gamma_j$, gave $\ln \gamma_i^\infty$ and $\ln \gamma_j^\infty$ at the intercepts, and finally $\gamma_i^\infty$ and $\gamma_j^\infty$ were obtained. The computed $\gamma_i^\infty$ and $\gamma_j^\infty$ for Ga-In at 623 K and Ga-Tl at 973 K are presented in \autoref{tab:gamma_dji}.

\begin{table}
\caption{Fitted parameters applied in modelling $G_M$ by CFM and the  $Er_is$.
	{\label{tab:cfm_data}}}
	\scalebox{0.9}{
\begin{tabular*}{\tblwidth}{@{} LCCCCCCCCC@{} }
\toprule
  Alloy & Temp. &  &  & {\Large \multirow{2}{*}{$\frac{g}{R T}$}} & {\Large \multirow{2}{*}{$\frac{\nu_{12}}{R T}$}} & {\Large \multirow{2}{*}{$\frac{\nu_{13}}{R T}$}} & {\Large \multirow{2}{*}{$\frac{\nu_{23}}{R T}$}} &  \multicolumn{2}{c}{$Er_t\text{ }(\%)$}  \\   \cline{9-10}
  $i-j$ & $T\text{ }(K)$ & $\mu$ & $\nu$  &  &  &  &   & $G_M$ & $a_is$ \\
\midrule
  Ga-In & 623 & 1 & 1 & -3.5 & 1.0 & -0.1 & 0.0 & 1.61 & 1.20 \\
  Ga-Tl & 973 & 1 & 1 & -1.3 & 2.2 & 2.3  & 1.6 & 2.65 & 3.06 \\
\bottomrule
\end{tabular*}	}	
\end{table}

\begin{table*}[width=.9\textwidth,cols=10,pos=h]
\caption{The values of $\gamma_i^\infty \text{, } \gamma_j^\infty \text{, } D_{ji} \text{ and } D_{ij}$.{\label{tab:gamma_dji}}}
\scalebox{1.1}{
\begin{tabular*}{\tblwidth}{ L@{\hskip 0.5cm}C@{\hskip 0.5cm}C@{\hskip 0.5cm}C@{\hskip 0.5cm}C@{\hskip 0.3cm}CCCCCCCCCC }
\toprule
   & Exp. Temp. &  &  &  &   &  &  \multicolumn{8}{c}{Values at other temperatures}  \\   \cline{8-15}
   Alloy & $T\text{ }(K)$ & \multicolumn{4}{c}{Values at experimental temperatures} &  & \multicolumn{2}{c}{1073 K} &  & \multicolumn{2}{c}{1173 K}  &  &  \multicolumn{2}{c}{1273 K}  \\   \cline{3-6} \cline{8-9} \cline{11-12} \cline{14-15}
  $i-j$ & \cite{pap11,pap12,pap15} & $\gamma_i^\infty$ & $\gamma_j^\infty$ & $D_{ji}$ & $D_{ij}$ &  &  $D_{ji}$ & $D_{ij}$ &  & $D_{ji}$ & $D_{ij}$ & & $D_{ji}$ & $D_{ij}$ \\
\midrule
  $\text{Ga-In}^a$ & 623 & 2.4476 & 2.4296 & 0.8817 & 0.9482 &  & 0.9295 & 0.9696 &  & 0.9353 & 0.9721 &  & 0.9403 & 0.9743\\
$\text{Ga-Tl}^a$ & 973 & 5.2070 & 6.1858 & 0.8385 & 0.8200 &  & 0.8524 & 0.8353 &  & 0.8641 & 0.8482 &  & 0.8741 & 0.8592\\
$\text{In-Tl}^b$ & 723 & 1.5120 & 1.9580 & 1.1518 & 0.7530 &  & 1.0999 & 0.8260 &  & 1.0910 & 0.8396 &  & 1.0836 & 0.8512\\
\bottomrule
\multicolumn{15}{l}{\footnotesize{$^a\text{ } \gamma_i^\infty \text{ and } \gamma_j^\infty$ were obtained in this work by applying the method in \cite{pap13}}}\\
\multicolumn{15}{l}{\footnotesize{$^b\text{ } \gamma_i^\infty \text{ and } \gamma_j^\infty$ were obtained from \cite{pap15}}}\\
\end{tabular*}	}	
\end{table*}

For preliminary validations, the computed $\gamma_i^\infty$ and $\gamma_j^\infty$ were applied in MIVM to predict the thermodynamic activities of Ga and In in Ga-In at 623 K, and the thermodynamic activities of Ga and Tl in Ga-Tl at 973 K. The $Er_{a_is}$ values for the MIVM predictions against experimental data are $\pm 1.15\%$ and $\pm 2.55\%$ for Ga-In and Ga-Tl, respectively.

\subsection{Thermodynamic activities}
\label{results:activities}
The thermodynamic activities of all components in Ga-In-Tl ($a_{Ga}$, $a_{In}$ and $a_{Tl}$), which were predicted at 1073, 1173 and 1273 K, are presented in Tables \ref{tab:aGa} and \ref{tab:aGaaIn}. \autoref{tab:aGa} shows computed and experimental $a_{Ga}$ at 1073, 1173 and 1273 K, as well as the errors $(Er_{a_{Ga}} \text{ and } Er_{a_{Ga}}^\ast)$ for the two data set. The agreement between experimental and computed $a_{Ga}$, quantified by $Er_{a_{Ga}} \text{ and } Er_{a_{Ga}}^\ast$ estimations, was observed to be best at 1073 K systems $(\pm 8.07\%\text{; }\pm 0.0290)$, followed by 1173 K systems $(\pm 10.62\%\text{; }\pm 0.0338)$, then 1273 K systems $(\pm 14.32\%\text{; }\pm 0.0434)$. The errors were also observed to decrease as Ga-content increases. Mean absolute percentage and standard errors for the three temperatures (combined) are $\pm 11.00\%\text{ and }\pm 0.0359$, respectively, and that is satisfactory for further predictions. The non-existing experimental activities of In and Tl at 1073, 1173 and 1273 K have been predicted based on the satisfactory agreements between computed and experimental $a_{Ga}$ at the three temperatures.

\begin{figure*}[width=1.0\textwidth,cols=4,pos=!htbp]
\centering
\includegraphics[width=\textwidth]{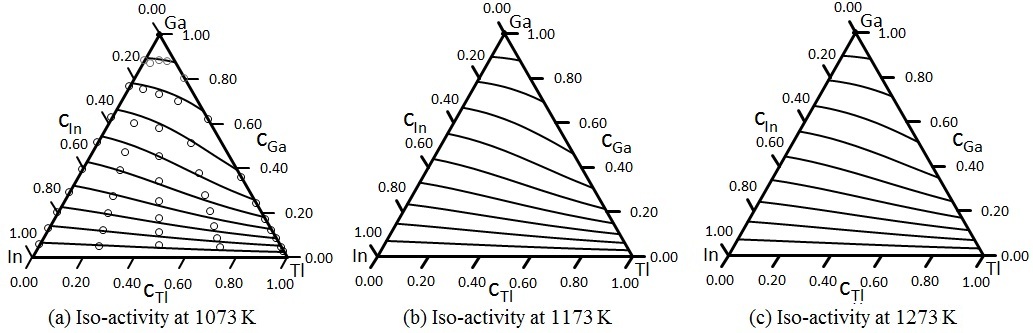}
\caption{Nine iso-activity curves $0.1-0.9$ (from bottom to top) of Ga in Ga-In-Tl system in the temperature range $1073-1273$ K. Circle represents experimental iso-activity \cite{pap01}.}
\label{fig:isoGa}
\includegraphics[width=\textwidth]{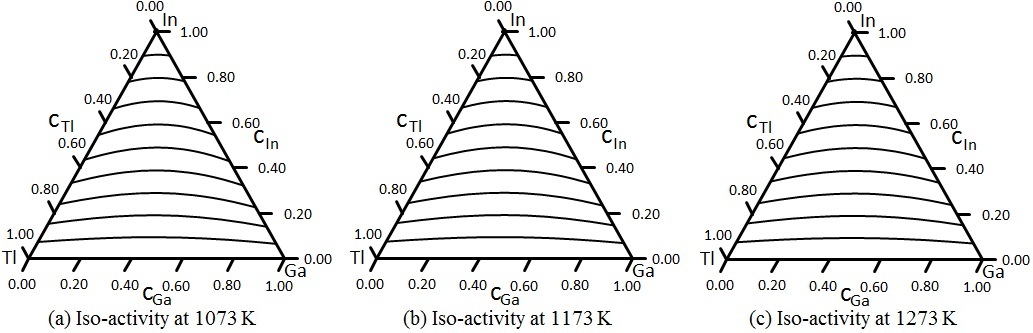}
\caption{Nine iso-activity curves $0.1-0.9$ (from bottom to top) of In in Ga-In-Tl system in the temperature range $1073-1273$ K.}
\label{fig:isoIn}
\includegraphics[width=\textwidth]{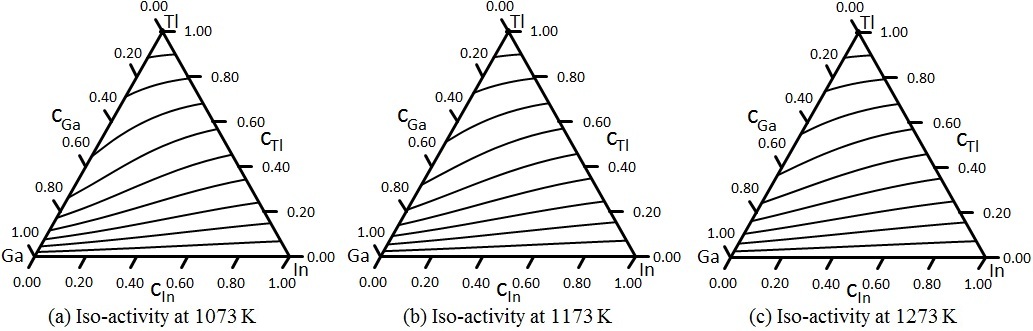}
\caption{Nine iso-activity curves $0.1-0.9$ (from bottom to top) of Tl in Ga-In-Tl system in the temperature range $1073-1273$ K.}
\label{fig:isoTl}
\end{figure*}

Iso-activity curves for Ga in Ga-In-Tl at $1073-1273$ K are presented in \autoref{fig:isoGa}. \autoref{fig:isoGa} (a $-$ c) show that $a_{Ga}$ exhibits positive deviation from Raoultian behaviour in all compositions and temperatures, however, Ga portrays ideal behaviour when $c_{Ga} > 0.9$. The $a_{Ga}$ increases in value from the Ga-In binary section to the Ga-Tl binary section. Also, $a_{Ga}$ increases with the addition of Tl within the ternary sections. Therefore, the addition of Tl reduces the tendency for compound formation, which implies a reduction in the chemical short range order between the atoms of Ga and In in Ga-In system. As temperature increases, the $a_{Ga}$ reduces $-$ approaching Raoult’s law. The iso-activity curves for Ga in Ga-In-Tl at 1073 K have reasonable agreement with the reported experimental iso-activity curve of Katayama et al. \cite{pap01}, this is shown in \autoref{fig:isoGa} (a).

\autoref{fig:isoIn} (a $-$ c) show the iso-activity curves for In in Ga-In-Tl at $1073-1273$ K, which reveals that $a_{In}$ possesses slight positive deviation from ideal values. The least deviations are observed where Ga and Tl atoms have equal concentrations. This means that In atoms have reducing tendencies of segregation in the Ga-In-Tl system as Ga and Tl approach equiatomic states. The deviations are largest at the binary sections (Ga-In, followed by In-Tl), and they reduce as temperature increases.

Iso-activity curves for Tl in Ga-In-Tl at $1073-1273$ K are presented in \autoref{fig:isoTl} (a $-$ c). The $a_{Tl}$ in the Ga-In-Tl systems exhibit positive deviation from ideality in all compositions and temperatures, excluding $c_{Tl}>0.9$ where $a_{Tl}$ is ideal. The $a_{Tl}$ in the Ga-Tl binary section reduces (approaches Raoultian behaviour) with the gradual addition of In atoms, all through the Ga-In-Tl ternary sections, until the In-Tl binary section. This implies that Tl atoms have a higher tendency of forming hetero-coordinated mixtures in In-Tl systems than in Ga-Tl systems. As expected, the deviations of $a_{Tl}$ from ideality reduces with temperature increase.

\begin{table}[!hp]
\caption{Comparison of the predicted values of activity of Ga in the Ga-In-Tl system with experimental data at $1073 - 1273$ K.{\label{tab:aGa}}}
\scalebox{.75}{
\begin{tabular}{ C@{\hskip 0.3cm}C@{\hskip 0.3cm}C@{\hskip 0.6cm}C@{\hskip 0.3cm}C@{\hskip 0.3cm}C@{\hskip 0.3cm}C@{\hskip 0.3cm}C@{\hskip 0.3cm}C@{\hskip 0.3cm}C }
\toprule
   &  &  &  & $a_{Ga}$ &  & &  & $a_{Ga}$ &  \\ 

   &  &  &  \multicolumn{3}{c}{MIVM}  & & \multicolumn{3}{c}{Exp. \cite{pap07}}  \\   \cline{4-6}  \cline{8-10}
  $c_{Ga}$ & $c_{In}$ & $c_{Tl}$ & 1073 K & 1173 K & 1273 K &  & 1073 K & 1173 K & 1273 K \\ 
\midrule
0.100 & 0.225 & 0.675 & 0.2647 & 0.2406 & 0.2219 & & 0.221 & 0.174 & 0.142\\
0.200 & 0.200 & 0.600 & 0.4344 & 0.4026 & 0.3776 & & 0.396 & 0.359 & 0.329\\
0.400 & 0.150 & 0.450 & 0.6253 & 0.5985 & 0.5767 & & 0.643 & 0.590 & 0.549\\
0.600 & 0.100 & 0.300 & 0.7368 & 0.7220 & 0.7097 & & 0.722 & 0.664 & 0.618\\
0.800 & 0.050 & 0.150 & 0.8441 & 0.8396 & 0.8359 & & 0.872 & 0.832 & 0.800\\
\\
0.100 & 0.450 & 0.450 & 0.2109 & 0.1956 & 0.1836 & & 0.147 & 0.137 & 0.129\\
0.200 & 0.400 & 0.400 & 0.3617 & 0.3407 & 0.3240 & & 0.366 & 0.337 & 0.315\\
0.400 & 0.300 & 0.300 & 0.5609 & 0.5422 & 0.5269 & & 0.542 & 0.523 & 0.508\\
0.600 & 0.200 & 0.200 & 0.6996 & 0.6889 & 0.6800 & & 0.723 & 0.705 & 0.689\\
0.800 & 0.100 & 0.100 & 0.8323 & 0.8290 & 0.8262 & & 0.843 & 0.828 & 0.816\\
\\
0.100 & 0.675 & 0.225 & 0.1721 & 0.1627 & 0.1553 & & 0.147 & 0.122 & 0.105\\
0.200 & 0.600 & 0.200 & 0.3073 & 0.2941 & 0.2834 & & 0.314 & 0.314 & 0.313\\
0.400 & 0.450 & 0.150 & 0.5103 & 0.4977 & 0.4874 & & 0.533 & 0.529 & 0.525\\
0.600 & 0.300 & 0.100 & 0.6696 & 0.6621 & 0.6559 & & 0.699 & 0.675 & 0.655\\
0.760 & 0.180 & 0.060 & 0.7911 & 0.7879 & 0.7852 & & 0.818 & 0.786 & 0.760\\
\bottomrule
\multicolumn{10}{l}{$Er_{a_{Ga}}=\pm 11.00\%\text{; }Er_{a_{Ga}}^\ast=\pm 0.0359$}\\
\end{tabular}
}	
\end{table}

\begin{table}
\caption{Prediction of activities of In and Tl in the Ga-In-Tl system at $1073 - 1273$ K. {\label{tab:aGaaIn}}}
\scalebox{0.75}{
\begin{tabular}{ C@{\hskip 0.3cm}C@{\hskip 0.3cm}C@{\hskip 0.6cm}C@{\hskip 0.3cm}C@{\hskip 0.3cm}C@{\hskip 0.3cm}C@{\hskip 0.3cm}C@{\hskip 0.3cm}C@{\hskip 0.3cm}C }
\toprule
   &  &  &  & $a_{Ga}$ &  & &  & $a_{Ga}$ &  \\ 
   &  &  &  \multicolumn{3}{c}{MIVM}  & & \multicolumn{3}{c}{Exp. \cite{pap07}}  \\   \cline{4-6}  \cline{8-10}
  $c_{Ga}$ & $c_{In}$ & $c_{Tl}$ & 1073 K & 1173 K & 1273 K &  & 1073 K & 1173 K & 1273 K \\ 
\midrule
0.100 & 0.225 & 0.675 & 0.2585 & 0.2558 & 0.2535 & & 0.7148 & 0.7116 & 0.7088\\
0.200 & 0.200 & 0.600 & 0.2179 & 0.2165 & 0.2153 & & 0.6775 & 0.6705 & 0.6645\\
0.400 & 0.150 & 0.450 & 0.1565 & 0.1559 & 0.1553 & & 0.6222 & 0.6041 & 0.5891\\
0.600 & 0.100 & 0.300 & 0.1090 & 0.1080 & 0.1071 & & 0.5645 & 0.5321 & 0.5060\\
0.800 & 0.050 & 0.150 & 0.0626 & 0.0611 & 0.0599 & & 0.4330 & 0.3914 & 0.3592\\
\\
0.100 & 0.450 & 0.450 & 0.4806 & 0.4778 & 0.4755 & & 0.5189 & 0.5133 & 0.5083\\
0.200 & 0.400 & 0.400 & 0.4158 & 0.4144 & 0.4131 & & 0.5003 & 0.4910 & 0.4831\\
0.400 & 0.300 & 0.300 & 0.3112 & 0.3099 & 0.3088 & & 0.4678 & 0.4494 & 0.4342\\
0.600 & 0.200 & 0.200 & 0.2220 & 0.2194 & 0.2173 & & 0.4217 & 0.3933 & 0.3706\\
0.800 & 0.100 & 0.100 & 0.1280 & 0.1247 & 0.1219 & & 0.3118 & 0.2797 & 0.2551\\
\\
0.100 & 0.675 & 0.225 & 0.6857 & 0.6846 & 0.6836 & & 0.3003 & 0.2932 & 0.2871\\
0.200 & 0.600 & 0.200 & 0.6076 & 0.6067 & 0.6059 & & 0.2919 & 0.2825 & 0.2746\\
0.400 & 0.450 & 0.150 & 0.4708 & 0.4684 & 0.4664 & & 0.2731 & 0.2585 & 0.2467\\
0.600 & 0.300 & 0.100 & 0.3419 & 0.3371 & 0.3332 & & 0.2406 & 0.2215 & 0.2065\\
0.760 & 0.180 & 0.060 & 0.2285 & 0.2228 & 0.2182 & & 0.1887 & 0.1691 & 0.1542\\
\bottomrule
\end{tabular}
}	
\end{table}
    
\section{Conclusions}
\label{conclusions}

The thermodynamic activities of all components in the Ga-In-Tl system have been successfully studied with the aid of MIVM \cite{pap02}. This implies that the $\gamma_{Ga}^\infty$ and $\gamma_{In}^\infty$ computed for Ga-In and the $\gamma_{Ga}^\infty$ and $\gamma_{Tl}^\infty$ for Ga-Tl liquid binary alloys are reliable, and the method for computing $\gamma_{i}^\infty$ and $\gamma_{j}^\infty$ of binary liquid alloys using CFM \cite{pap13} is credible.

Following a satisfactory agreement between experimental and computed activities of Ga, especially at 1073 K, activities of Ga, In and Tl in Ga-In-Tl were calculated across all composition range and summarised with iso-activity plots for the three components. The activities of Ga, In and Tl exhibits positive deviation from Raoult’s law, and this reduces with temperature increase, in all compositions. It is worthy to note that the addition of Tl increases the tendency of segregation of Ga in the Ga-In-Tl system. Also, the activity of In approaches ideality when Ga and Tl are about equiatomic.

From this work, $\gamma_{i}^\infty$ and $\gamma_{j}^\infty$ values for Ga-In and Ga-Tl liquid alloys have been obtained, and this has contributed to the knowledge of the thermodynamics of Ga-In-Tl alloys, which eventually, is a contribution to the thermodynamic database of lead-free solders, which is important for the calculation of phase equilibria and prediction of several alloy properties.

\bibliographystyle{elsarticle-num-names}

\bibliography{GaXGaInTl}

\end{document}